\documentclass[aps,pre,showpacs,twocolumn,superscriptaddress]{revtex4-1}
\usepackage{graphicx,color,epsfig}
\usepackage{amssymb,amsmath}

\begin{document}
\title{Chimeras and clusters emerging from robust-chaos dynamics}
\author{M. G. Cosenza}
\affiliation{School of Physical Sciences \& Nanotechnology, Universidad Yachay Tech, Urcuqu\'i,  Ecuador}
\affiliation{Universidad de Los Andes, M\'erida, Venezuela}
\author{O. Alvarez-Llamoza}
\affiliation{Grupo de Simulaci\'on, Modelado, An\'alisis y Accesabilidad, Universidad Cat\'olica de Cuenca, Cuenca, Ecuador}
\author{A. V. Cano}
\affiliation{Institute for Integrative Biology, ETH, Zurich, Switzerland}
\affiliation{Swiss Institute of Bioinformatics, Lausanne, Switzerland}

\date{February 2021}

\begin{abstract}
	\section*{Abstract}
	We show that dynamical clustering, where a system segregates into distinguishable subsets of synchronized elements, and chimera states, where differentiated subsets of synchronized and desynchronized
	elements coexist, can emerge in networks of globally coupled robust-chaos oscillators. 
We describe the collective behavior of a model of globally coupled robust-chaos maps in terms of statistical quantities, and characterize clusters, chimera states,
synchronization, and incoherence on the space of parameters of the system.
 We employ the analogy between the local dynamics of
a system of globally coupled maps with the response dynamics of a
single driven map.  We
interpret the 
occurrence of clusters and chimeras in a globally coupled system of robust-chaos maps in terms of windows of
periodicity and multistability induced by a drive on the local robust-chaos map.
Our results
show that robust-chaos dynamics does not limit the formation of cluster and chimera states in networks
of coupled systems, as it had been previously conjectured.

\end{abstract}

\maketitle

\section{Introduction}
Many smooth nonlinear dynamical systems possess chaotic attractors embedded with a
dense set of periodic orbits for any range of parameter values.
Therefore, in practical systems operating in chaotic mode, a slight perturbation of a parameter may drive
the system out of chaos. Alternatively, there exist dynamical
systems that exhibit the property of robust chaos \cite{Yorke,Kawabe,Priel,Potapov,Sprott,Gallas}. 
A chaotic attractor is said to be robust if, for its
parameter values, there exists a neighborhood in the parameter space where
windows of periodic orbits are absent and the chaotic attractor is unique \cite{Yorke}.

Robust chaos constitutes an advantageous feature in applications that require 
reliable functioning in a chaotic regime, in the sense that the chaotic behavior cannot be
removed by arbitrarily small fluctuations of the system parameters. 
For instance, networks of coupled maps with
robust chaos have been efficiently used in
communication and encryption algorithms \cite{Garcia} and they have been investigated for information transfer across scales in complex systems \cite{Cisneros}.
In addition, the existence of robust chaos allows for 
heterogeneity in the local parameters of a system of coupled oscillators,
while guaranteeing the performing of all the
oscillators in a chaotic mode.

On the other hand, systems possessing robust chaos may present limitations in the 
types of collective behaviors that they can achieve, in comparison with systems displaying periodic windows.  
For example, it has been conjectured
that the phenomenon of dynamical
clustering in globally coupled networks (where the system segregates into distinguishable
subsets of synchronized elements)
is only found when stable periodic windows are
present in the local elements \cite{CP,Manrubia,French}.
Recently, it has also been argued that
chimera states (i.e., coexistence of subsets of oscillators with synchronous 
and asynchronous dynamics) cannot emerge in networks of coupled oscillators
having robust chaotic attractors \cite{Semenova,Scholl}. 
 
The phenomenon of dynamical clustering is relevant as
it can provide a simple mechanism for the emergence of differentiation,
segregation, and ordering of elements in many physical and
biological systems \cite{Kaneko,Kaneko2}. Clustering
has been found in systems of globally coupled R\"ossler
oscillators \cite{Zanette}, neural networks \cite{Zanette2}, biochemical reactions \cite{Furusawa},
and has been observed experimentally in arrays of globally coupled electrochemical oscillators \cite{Wang}
and globally coupled salt-water oscillators \cite{Yamada}.
In addition, the study of chimera states currently attracts
much interest; for reviews see, \cite{Scholl2,Panaggio}. 
Chimera states have been found in networks of nonlocally coupled phase oscillators \cite{Kuramoto,Abrams},
in systems with local \cite{Laing1,Clerc,Bera,Hiz} 
and global \cite{Sen,Pik,Schmidt,Mis,We,Cano} interactions, 
and in networks of time-discrete maps \cite{Giulia,Omel2,Gupte,Ryba}.
These states have been investigated in a diversity of contexts 
\cite{Ulo,Kanas,Bastidas,Dutta,Semenov,Lima,Roth,JC,Fila}.  
Chimera states have been observed in experimental settings, such as 
populations of chemical oscillators \cite{Showalter}, coupled lasers \cite{Hart}, 
optical light modulators \cite{Roy}, 
electronic \cite{Larger}, and mechanical \cite{Martens,Blaha}  
oscillator systems. 
It has been shown that clustering is closely related to the formation of chimera states 
in systems of globally coupled periodic oscillators \cite{Schmidt}.

In this paper, we investigate the occurrence of dynamical clustering and chimera states in systems of coupled robust-chaos oscillators. In Sec.~II, we describe the characterization of synchronization, cluster and chimera states in globally coupled systems. 
In Sec.~III we consider a network of globally coupled robust-chaos maps and show that cluster and chimera states can actually emerge in this system for several values of parameters. 
In Sec.~IV, we employ the analogy between the local dynamics of
the globally coupled system with the response dynamics of a
single driven map. We 
interpret the 
occurrence of clusters and chimeras in the globally coupled system in terms of windows of
periodicity induced by the drive on the local robust-chaos map. Conclusions are presented in Sec~V.

\section{Methods}
A global interaction in a system can be described as a field or influence
acting on all the elements in the system. As a simple model of an autonomous dynamical
system subject to a global interaction, we consider a system of $N$ maps coupled in the form
\begin{equation}
\label{Syst}
{x}^i_{t+1}=(1-\epsilon) f({x}^i_t) + \epsilon h_t(x_t^j | j \in S),
\end{equation}
where $x^i_t$ $(i = 1,2,\ldots,N)$ describes the state variable of the
$i$th map in the system at discrete time $t$, the function $f$ expresses the local dynamics of the maps, the function
$h_t$ represents a global field
that depends on the states of the elements in a given subset $S$ of the system, at time $t$,
and the parameter $\epsilon$ measures the strength of the coupling of the maps to the field. 
The form of the coupling in Eq.~(\ref{Syst}) is assumed in the commonly used diffusive form.
The function $h_t$ may not depend on all the elements, but it must be shared by all the elements of the system to be a global interaction.

A collective state of synchronization or coherence takes place in the system Eq.~(\ref{Syst}) when 
$x^i_t=x^j_t$,  $\forall i,j$ for asymptotic times.  
A desynchronized or incoherent state  corresponds to $x^i_t \neq x^j_t$ 
$\forall i,j$ for all times.
Dynamical clustering occurs when the
system segregates into a number of $K$ distinct clusters or subsets of elements  
such that elements in given subset are synchronized among themselves. In other words,
$x^i_t=x^j_t=X^\xi_t$, $\forall i,j$ in
the $\xi$th cluster, where $X^\xi_t$ denotes the value of $x^i_t$ in that cluster,
with $\xi=1,\ldots,K$.
If $n_\xi$ is the number of elements belonging to the $\xi$th cluster,
then its relative size is $p_\xi=n_\xi/N$. 
In general, the number of clusters, their size, and their dynamical evolution (periodic, quasiperiodic, or chaotic) depend on the initial conditions and parameters of the system. 
A chimera state consists of 
the coexistence of one or more clusters and a subset of desynchronized elements.
If there are $K$ clusters, the fraction of elements in the system belonging to clusters is 
$p=\sum_{\xi=1}^K  n_\xi /N$ while 
$(1-p)$ is the fraction of elements in the desynchronized subset. 
 
In practical applications, we consider that two elements $i$ and $j$ belong to a cluster at time $t$ 
if the distance between their state variables,
defined as
\begin{equation}
 d_{ij}(t)=|x_t^i-x_t^j|,
\end{equation}
is less than a threshold value $\delta$, i.e., if $d_{ij} < \delta$. The choice of $\delta$ should be appropriate for 
achieving differentiation between closely evolving clusters. 
Then, we calculate the fraction of elements that belong to some cluster at time $t$ as \cite{Zanette}
\begin{equation}
p(t)= 1-\frac{1}{N}\sum_{i=1}^N \prod_{j=1, j\neq i}^N \Theta[d_{ij}(t)-\delta],
\end{equation}
where $\Theta(x)=0$ for $x<0$ and $\Theta(x)=1$ for $x\geq 0$. We refer to $p$ as the asymptotic time-average of $p(t)$.
Then, a clustered state in the system can be characterized by the value $p=1$, while an incoherent state in the system corresponds to $p \to 0$.
The values $p_{\mbox{\scriptsize min}} < p < 1$ characterize a chimera state, 
where $p_{\mbox{\scriptsize min}}$ is the minimum cluster size to be taken into consideration. 

A synchronization state corresponds to the presence of a single cluster of size $N$ and  has also the value $p=1$. 
To distinguish a synchronization state from a multicluster state, we calculate the asymptotic time-average
$\langle \sigma \rangle$ as
\begin{equation}
\langle \sigma \rangle = \frac{1}{T-\tau}\sum_{t=\tau}^{T} \sigma_t,
\end{equation}
where $\tau$ is the number of discarded transients, $T$ is a sufficiently large time, and $\sigma_t$ is
the instantaneous standard deviation of the distribution of state variables defined by
\begin{equation}
\sigma_t  = \left[  \frac{1}{N} \sum_{i=1}^N   (x_t^i- \bar x_t)^2  \right]^{1/2},
\end{equation}
where
\begin{equation}
\bar{x}_t=\frac{1}{N} \sum_{j=1}^N x^j_t.
\end{equation}

Statistically, a synchronization state is characterized by the values $\langle \sigma \rangle=0$ and $p=1$,  
while a cluster state corresponds to $\langle \sigma \rangle>0$ and $p=1$. 
Chimera states are characterized by $\langle \sigma \rangle>0$ and $p_{\mbox{\scriptsize min}} < p < 1$,
and desynchronization is described by $\langle \sigma \rangle>0$, $p < p_{\mbox{\scriptsize min}}$.
In this paper we set $\delta=10^{-6}$ and $p_{\mbox{\scriptsize min}}=0.05$. 

Note that, in systems with local or long-range interactions where there is a natural spatial ordering, the synchronized and desynchronized domains for chimera states are localized in space. In contrast, 
globally coupled systems lack the notion of spatial order. Thus, the of chimera and cluster states in our system are characterized in terms of the statistical quantities $\langle \sigma \rangle$ and $p$, not on the spatial location of synchronized and desynchronized domains. 

\section{Results and Discussion}

\subsection*{Chimeras and clusters in globally coupled robust chaos maps}
Let us consider a network of globally coupled maps described by the equations
\cite{Kaneko}
\begin{equation}
\label{SystM}
{x}^i_{t+1}=(1-\epsilon) f({x}^i_t) + \frac{\epsilon}{N} \sum_{j=1}^N f({x}^j_t), 
\end{equation}
where the global interaction function is the mean field of the system,
\begin{equation}
h_t=\frac{1}{N} \sum_{j=1}^N f({x}^j_t).
\label{meanF} 
\end{equation}

\begin{figure}[h]
	\includegraphics[scale=0.62]{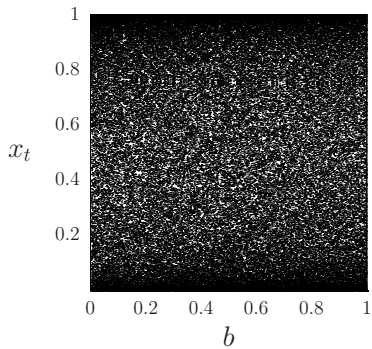}
	\caption{Bifurcation diagram of the map Eq.~(\ref{Ali}) as a function of the parameter $b$.}
	\label{FAli}
\end{figure}

As local dynamics exhibiting robust chaos, we consider the following smooth, unimodal map defined
on the interval $x \in [0,1]$ \cite{Andrecut},
\begin{equation}
\label{Ali}
x_{t+1}=f(x_t)= \frac{1-b^{(1-x_t)x_t}}{1-b^{1/4}}.
\end{equation}
which is chaotic with no periodic windows on the parameter interval $b \in [0,1]$.
On this interval, the Lyapunov exponent of map Eq.~(\ref{Ali}) has the constant value $\lambda=\ln 2$. 
The bifurcation diagram of map Eq.~(\ref{Ali}) in Figure~\ref{FAli} shows the absence of periodicity
in the interval $b \in [0,1]$.

\begin{figure}[h]
	\includegraphics[scale=.27]{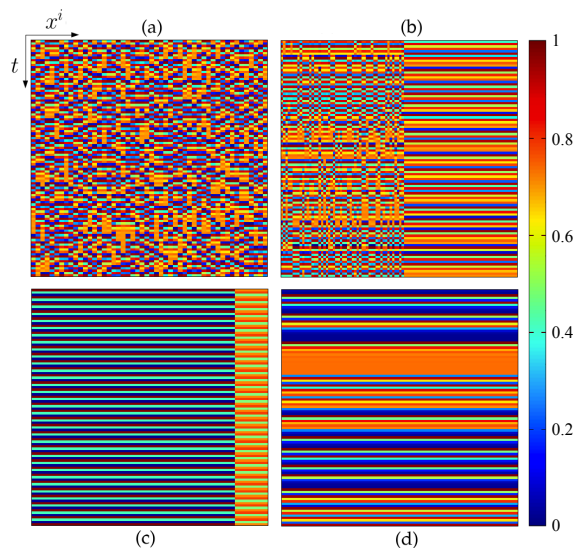}
	\caption{Asymptotic evolution of the states $x^i$ (horizontal axis) as a function of time (vertical axis) for the system
		Eqs.~(\ref{SystM})  and(\ref{Ali})
		with size $N=100$ and fixed $b=0.5$, for different values of the coupling parameter.
		Random initial conditions are uniformly distributed in the interval $[0,1]$. 
		After discarding $10^4$ transients, $100$ iterates $t$ are displayed. Ordering of the map indexes is explaining in the text. Color code: two elements $i,j$ 
share the same color if $x^i_t= x^j_t$.
		(a) Incoherent or desynchronized state, $\epsilon= 0.15$.
		(b) Chimera state, $\epsilon=0.2$. 
		(c) Two-cluster state,  $\epsilon= 0.39$. 
		(d) Synchronization,  $\epsilon=0.6$.}
	\label{f1}
\end{figure}

Figure~\ref{f1} shows the asymptotic temporal evolution of the states of the system Eqs.~(\ref{SystM}) and (\ref{Ali}), 
for different values of parameters. 
Since the system is globally coupled, there is no natural spatial ordering. For visualization purposes, 
the indexes $i$ are ordered at time $t=10^4$ such that $i<j$ if $x^i_t < x^j_t$ and kept fixed afterwards.
The values of the states $x^i_t$ are represented by distinct color coding; two elements $i,j$ 
share the same color if $x^i_t= x^j_t$. 
A desynchronized state is displayed in Fig.~\ref{f1}(a) and a complete synchronization state occurs in Fig.~\ref{f1}(d), while
a chimera state and a two-cluster state are visualized in Figs.~\ref{f1}(b) and \ref{f1}(c), respectively.

\begin{figure}[h]
\includegraphics[scale=.25,angle=90]{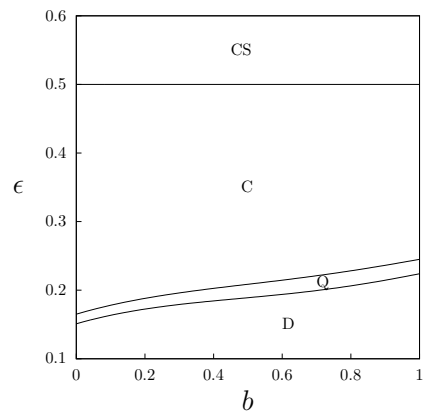}
\caption{Phase diagram on the plane $(\epsilon,b)$ for the autonomous system Eqs.~(\ref{SystM})
and (\ref{Ali}) with size $N=500$. 
For each data point, the quantities $p$ and $\langle \sigma \rangle$ are obtained by 
averaging over $50$ realizations of random
initial conditions $x^i_0$ uniformly distributed in the interval $[0,1]$. 
Labels indicate different collective states.
CS: synchronization; C: cluster states; Q: chimera states; D: desynchronization.}
\label{f2}
\end{figure}

Figure~\ref{f2} shows the collective states arising in 
the system Eqs.~(\ref{SystM}) and (\ref{Ali}) on the space of parameters $(\epsilon,b)$, characterized through the quantities 
$p$ and $\langle \sigma \rangle$. Labels indicate the regions where these behaviors occur:
CS: complete synchronization; C: cluster states; Q: chimera states, and D: desynchronization.

The linear stability analysis for the complete synchronization state in the globally coupled system Eq.~(\ref{SystM}) 
shows that this state is stable if the following condition is satisfied \cite{Kaneko},
\begin{equation}
 |(1-\epsilon)e^\lambda| < 1,
\end{equation}
where $\lambda$ is the Lyapunov exponent for the local map $f(x)$. For the map Eq.~(\ref{Ali}), we obtain that the
completely synchronized state is stable for $1/2 <\epsilon <3/2$, which agrees with the numerical characterization for this state performed in Fig.~\ref{f2}. 
Figure~\ref{f2} reveals that both cluster and chimera states can arise in globally coupled map networks
for appropriate values of parameters,
even when the individual maps lack periodic windows. 
Clusters and chimera states regions occur adjacent to each other for an intermediate range of values of the 
coupling parameter $\epsilon$ 
on the phase diagram of  Fig.~\ref{f2}.   
In fact, chimeras and clusters are 
closely related collective states 
in systems subject to global interactions \cite{Pik}.
Chimera states appear to mediate between dynamical clustering and incoherence.

\begin{figure}[h]
	\includegraphics[scale=0.28]{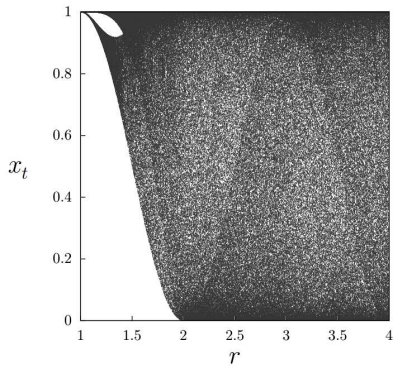}
	\caption{Bifurcation diagram of the map Eq.~(\ref{Aguirre}) as a function of the parameter $r$.}
	\label{Fagui}
\end{figure}

Multicluster chimera states are also possible in 
systems of globally coupled robust chaos maps.
As an illustration, consider the smooth unimodal map \cite{Aguirre}, 
\begin{equation}
\label{Aguirre}
f(x_t)= \sin^2({r\arcsin({\sqrt{x_t}}})),
\end{equation}
defined on the interval $x_t \in [0,1]$ for parameter values
$r > 1$.  Figure~(\ref{Fagui}) shows the bifurcation diagram of the iterates of map Eq.~(\ref{Aguirre}) as
a function of the parameter $r$.
The dynamics of the map displays robust chaos with no periodic windows for $r > 1$. The Lyapunov
exponent is $\lambda = \ln r$ \cite{Aguirre}.

\begin{figure}[h]
\includegraphics[scale=0.32]{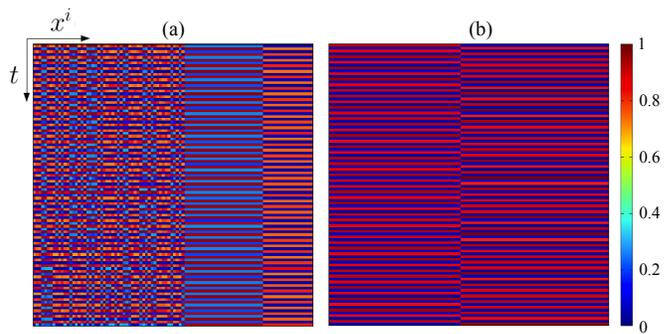}
\caption{Asymptotic states $x^i$ (horizontal axis) as a function of time (vertical axis) for the system Eqs.~(\ref{SystM}) with size $N=100$ and local map
 Eq.~(\ref{Aguirre}), for different values of parameters.
Initial conditions and ordering of the maps are similar to those in Fig.~\ref{f1}.  Color code: two elements $i,j$ 
share the same color if $x^i_t= x^j_t$.
 (a) Chimera state with two clusters, $r=3$, $\epsilon=0.235$.
 (b) Two-cluster state, $r=3$, $\epsilon= 0.272$.}
 \label{f0}
\end{figure}

Figure~(\ref{f0}) shows the temporal evolution of the states of the globally coupled 
system Eqs.~(\ref{SystM}) with the local map Eq.~(\ref{Aguirre}),
for different values of parameters. 
A chimera state with multiple clusters occurs in Fig.~\ref{f0}(a), while
a two-cluster state is shown in Fig.~\ref{f0}(b).
Multichimera states or multiheaded chimeras (coexistence of multiple localized domains of incoherence and coherence) have been reported in systems with long-range interactions \cite{Omelchenko3}. However, those states are not equivalent to a chimera state with multiple clusters in a globally coupled system, such as Fig.~\ref{f0}(a), where there is no notion of locality.

\subsection*{Dynamics of clusters and chimera states with global interactions}
Consider a chimera state consisting of $K$ clusters and a desynchronized subset
in the system of globally coupled maps Eq.~(\ref{Syst}). 
The dynamics of this state can be described by the equations
 \begin{equation}
  \label{quimera}
 \begin{array}{cc}
   X^\xi_{t+1}= (1-\epsilon) f(X^\xi_t) + \epsilon h_t, & \xi=1,\ldots,K, \\
 x^j_{t+1}= (1-\epsilon) f(x^j_t) + \epsilon h_t, & j=1,\ldots,(1-p)N.
   \end{array}
 \end{equation}
The mean field Eq.~(\ref{meanF}) in  a chimera state can be expressed as
the sum of two contributions
\begin{equation}
h_t= h_C +h_I,
\end{equation}
where 
 \begin{eqnarray}
 h_C &=&  \sum_{\xi=1}^K  p_\xi f(X^\xi_t), \\
h_I &=& \frac{1}{N}  \sum_{j=1}^{(1-p)N}  f(x^j_t).
\end{eqnarray}
The term $h_C$ is the contribution to the mean field corresponding to elements belonging to clusters, 
whereas $h_I$ is the average of the states of the
elements belonging to the incoherent subset.

\begin{figure}[h]
\includegraphics[scale=0.3]{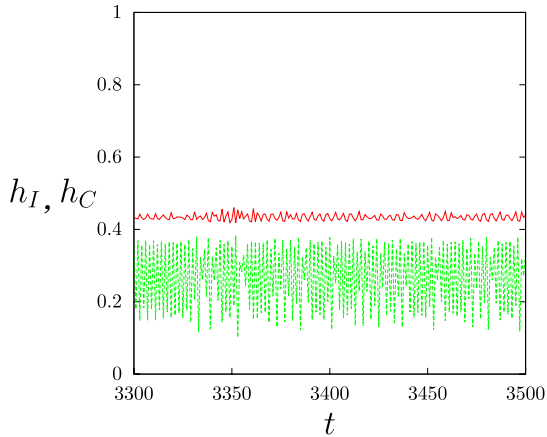}
\caption{Cluster $h_C$ (green line) and incoherent 
$h_I$ (red line) contributions to the mean field of the system Eqs.~(\ref{SystM}) and (\ref{Ali}), 
as functions of time. Fixed parameters $b=0.5$,
$\epsilon=0.19$, and size $N=10^4$.}
\label{f3}
\end{figure}

Figure~\ref{f3} shows the temporal behavior of both contributions $h_C$ and $h_I$ in a chimera
state for the globally coupled autonomous system Eqs.~(\ref{SystM}) and (\ref{Ali}).
The time evolution of the cluster contribution $h_C$
is chaotic, similar to that of the local map Eq.~(\ref{Ali}), but $h_C$ has a smaller amplitude. 
In general, the form of $h_C$ can be approximated as $h_C \approx Af(y_t)$, 
where $A<1$ represents a modulation factor reflecting the partition into several clusters.
On the other hand, Fig.~\ref{f3} reveals that the time series of $h_I$ fluctuates about a mean value 
with a small dispersion, 
corresponding to the superposition of the dynamics of many incoherent chaotic elements.
Thus, for the given parameter values, 
the incoherent contribution to the mean field for a large system size can be expressed approximately as a constant; i.~e., $h_I \approx k$. 

The dynamics of the globally coupled system Eqs.~(\ref{Syst}), where
each map is subject to a feedback field $h_t$,
can be compared to that of a replica system of maps subject to a global external drive
$g(y_t)$ in the form
 \begin{equation}
  \label{Edriven}
 \begin{array}{l}
x^i_{t+1}= (1-\epsilon) f(x^i_t) + \epsilon g(y_t), \\
y_{t+1} = g(y_t) .
   \end{array}
 \end{equation}

It has been shown that an analogy between the autonomous system Eq.~(\ref{Syst}) 
and the driven system Eq.~(\ref{Edriven})
can be established  when the time evolution of the field $h_t$ is identical to that of 
the function $g(y_t)$ \cite{CP}. 
Then, the drive-response dynamics at the local level in both systems
are similar, and therefore their corresponding emerging collective states can be 
equivalent for some appropriate parameter values and initial conditions.
In particular, chimera or cluster states in the system Eq.~(\ref{Edriven})
should be induced by an external drive function of the form 
$g(y_t)=A f(y_t)+ k$, with $A$, $k$ constants, that imitates the mean field $h_t$.
The realization of these states depends on the parameters $A$ and $k$ of the drive, 
and on the coupling strength $\epsilon$.

\begin{figure}[h]
	\includegraphics[scale=0.32]{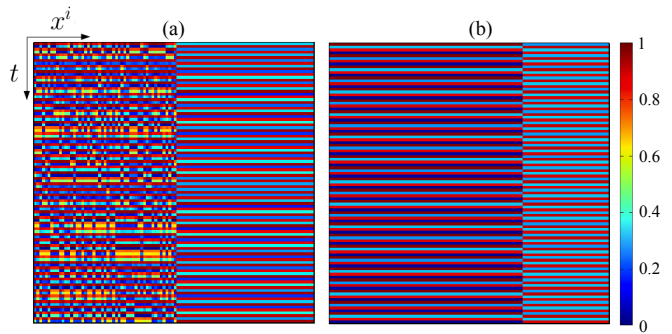}
	\caption{Asymptotic evolution of the states $x^i$ (horizontal axis) as a function of time (vertical axis) 
		for the driven system Eq.~(\ref{Edriven}) with size $N=100$
		and local map Eq.~(\ref{Ali}), for different values of the coupling parameter.
		Fixed values: $A=0.48$, $k=0.4$, $b=0.5$.  
		Random initial conditions $x_0^i$ are uniformly distributed in the interval $[0,1]$. 
		After discarding $10^4$ transients, $100$ iterates $t$ are displayed. Ordering of the maps is similar to that in Fig.~\ref{f1}.  Color code: two elements $i,j$ 
share the same color if $x^i_t= x^j_t$.
		(a) Chimera state;
		$\epsilon=0.198$.  
		(b) Two-cluster state; $\epsilon=0.272$.}
	\label{f4}
\end{figure}

Figure~\ref{f4} shows the temporal evolution of the states of the driven system 
Eqs.~(\ref{Edriven}) with local map Eq.~(\ref{Ali}),
for some values of parameters.  
A chimera state with a single cluster takes place in Fig.~\ref{f4}(a) for parameter
values $(\epsilon,b)$, where chimera states
also occur in the autonomous system Eqs.~(\ref{SystM}) and (\ref{Ali}), as seen in the corresponding phase diagram of Fig.~\ref{f2}. Figure~\ref{f4}(b) shows a two-cluster state for values $(\epsilon,b)$ located in the
region corresponding to clustered states in Fig.~\ref{f2}.
The dynamics of the driven system Eqs.~(\ref{Edriven}) displays multistability; depending on initial conditions, chimeras with different partitions may be induced for given parameters values $(\epsilon, b)$ in the region labeled Q in Fig.~\ref{f2}. Similarly, different initial conditions  may produce cluster states with different partition sizes for fixed parameter values in region C of Fig.~\ref{f2}.

The system Eqs.~(\ref{Edriven}) can be considered as 
$N$ realizations for different initial conditions of a single driven map
\begin{equation}
\label{Sdriven}
\begin{array}{ll}
x_{t+1}=& (1-\epsilon) f(x_t) + \epsilon g(y_t), \\
y_{t+1} =& g(y_t) .
\end{array}
\end{equation}
Analogously, each local map in the globally coupled system
Eqs.~(\ref{SystM})  
can be seen as subject to a field $h_t$
that eventually induces a collective state. 
Clustering in globally coupled systems of identical elements has been attributed to the
existence of periodic windows in the local dynamics \cite{Manrubia}. 
On the other hand, clustering is considered a prerequisite for the 
occurrence of chimera states in globally coupled systems \cite{Schmidt}.
Thus, to elucidate the origin of clusters and chimeras in system Eqs.~(\ref{SystM})
with local robust chaos, one can explore the response dynamics of the single driven map
Eq.~(\ref{Sdriven}) with a function of the form $g(y_t)=Af(y_t)+k$ and $f$ having robust chaos.
Then, if periodic windows are induced by the drive
on a single map,
one may expect that clusters and chimeras should
arise in a globally coupled system of those maps.

\begin{figure}[h]
	\includegraphics[scale=0.48]{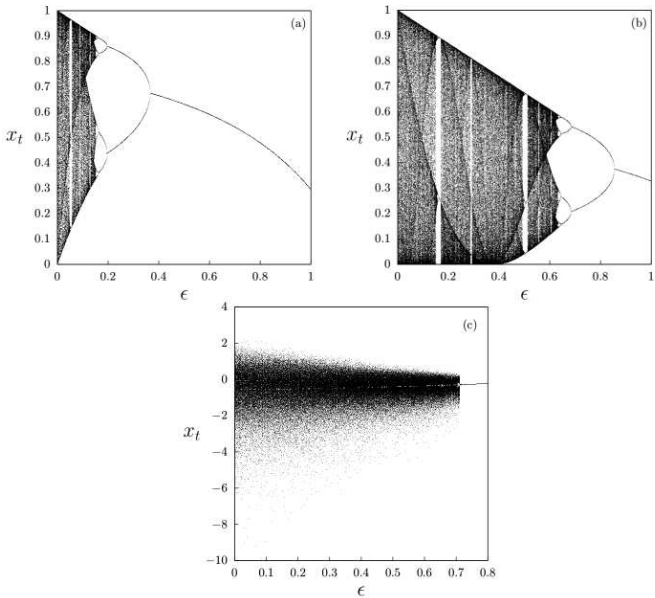}
	\caption{Bifurcation diagrams of the driven map $x_{t+1}=(1-\epsilon)f(x_t)$ in Eq.~(\ref{Sdriven})
		as a function of $\epsilon$ for different robust chaos maps $f$. 
		(a) $f(x)= \frac{1-b^{(1-x)x}}{1-b^{1/4}}$ with $b=0.5$.
		(b) $f(x)= \sin^2({r\arcsin({\sqrt{x}}}))$ with $r=3$. 
		(c)$f(x)= \ln |x|$.}
	\label{f7}
\end{figure}

Even a trivial function $g$ can modify the dynamics of a driven robust chaos map in Eq.~(\ref{Sdriven}) 
to produce periodic windows.
Figure~\ref{f7}(a) shows the bifurcation diagram of $x_t$ in Eq.~(\ref{Sdriven})
versus
$\epsilon$ for the map $f$ given by Eq.~(\ref{Ali})
with $g(y_t) \to 0$, which is equivalent to a rescaling of $f$.
Periodic windows typical of unimodal maps appear in the 
rescaled map $x_{t+1}=(1-\epsilon)f(x_t)$.
In general, the driven map 
Eq.~(\ref{Sdriven}) represents a rescaling of the robust-chaos map $f$
that acquires periodic windows. Similarly,
the periodic cluster states arising in the globally coupled system Eqs.~(\ref{SystM}) and (\ref{Ali})
are a consequence of the windows of periodicity induced locally by the mean field
$h_t$, in analogy to the periodic windows created by an external drive $g$ acting on a single map Eq.~(\ref{Ali}).
Different initial conditions may lead to different out-of-phase orbits with
diverse partitions 
that appear as clusters  
in the globally coupled system.
A synchronization state in the system Eqs.~(\ref{SystM}) and (\ref{Ali}) 
can be associated to the fixed point interval of the bifurcation diagram
of Fig.~\ref{f7}(a), while a desynchronization state
in the globally coupled system
is a manifestation of a chaotic regime as seen in Fig.~\ref{f7}(a). 
Nontrivial forms of the driving function can give rise to multistable behavior besides
periodic windows.
For example, we have verified that
a drive such as $g(y_t)=0.48f(y_t)+0.4$ in Eq.~(\ref{Sdriven}) 
induces a region of bistability between chaotic attractors
that expresses as chimera states in the associated globally coupled system Eqs.~(\ref{SystM}) and (\ref{Ali}).

These results suggest that
the emergence of cluster and chimera states in a globally coupled system of robust-chaos maps can be inferred
from the occurrence of
periodic windows in the 
response dynamics of a single map subject to an appropriate drive, as a function of parameters.
Figure~\ref{f7}(b) shows the corresponding bifurcation diagram of $x_{t+1}=(1-\epsilon)f(x_t)$ versus $\epsilon$ for the map $f$ given by Eq.~(\ref{Aguirre}) which also has robust chaos. Again, we see
the emergence of periodic windows as the coupling parameter is varied. A globally coupled system of these maps also shows
clusters and chimera states, as illustrated in Fig.~\ref{f0}.
Figure~\ref{f7}(c) presents the bifurcation diagram of 
$x_{t+1}=(1-\epsilon)f(x_t)$ versus $\epsilon$ for the 
logarithmic map $f=a+\ln|x|$, which possesses robust chaos on the parameter interval $a \in [-1,1]$ and its dynamics is unbounded \cite{Kawabe}. 
In contrast to Figs.~\ref{f7}(a)-(b), no periodic windows appear on the dynamics of 
the driven map Eq.~(\ref{Sdriven}) which remains unbounded; only chaotic orbits and a fixed point attractor appear. 
As a consequence, clusters and chimera states should not be expected in a globally coupled system of logarithmic maps.
In fact, only synchronization and nontrivial collective behavior have been observed in such a system \cite{Gallego}.
  
\section{Conclusions}
Networks of globally coupled identical oscillators are among the simplest symmetric spatiotemporal
systems that can display clustering and chimera behavior. 
Previous works have conjectured that these phenomena cannot occur
when the local oscillators possess robust-chaos attractors \cite{CP,Manrubia,French,Semenova,Scholl}. 
We have shown that the presence of global interactions can indeed allow for 
emergence of both cluster and chimera states in systems of coupled robust-chaos maps. Chimeras appear as partially ordered states between synchronization or clustering and incoherent behavior. 
We have found that chimera states are associated to
the formation of clusters in these systems, a feature
that has been observed in other globally coupled systems \cite{Schmidt}.

The existence of intrinsic periodic windows in the dynamics of local oscillators,
such as in logistic maps, is not essential for 
the emergence of clusters with periodic behavior
in a globally coupled system of those oscillators. 
Windows of periodicity and multistability can be induced in the dynamical response of
a robust-chaos map subject to an appropriate external forcing. Because of the analogy between a single driven map and the local dynamics of a globally coupled map system,
the global interaction field $h_t$ can also induce
periodic windows and multistability on local robust-chaos maps. Those are the essential ingredients for the occurrence of cluster and chimera states in globally coupled systems. Since clustering is a prerequisite for chimeras \cite{Schmidt}, a single driven robust-chaos map that develops periodic windows on some range of parameters allows us to infer that 
a globally coupled system of such maps shall also exhibit
cluster and chimera states on some range of parameters.
Conversely, a robust-chaos map,  such as the logarithmic or another singular map, that does not give rise to periodic windows when subject to a drive, implies that 
a system of globally coupled logarithmic or singular maps do not show clusters nor chimera states.

Further extensions of this work include the investigation
of chimera states in networks of globally coupled continuous-time dynamical systems possessing robust chaos or hyperbolic chaotic attractors, the study of interacting populations of robust-chaos elements, and the role of the range of interaction in a network of dynamical robust-chaos units.

\section*{Acknowledgment}
This work was supported by
Corporaci\'on Ecuatoriana para el Desarrollo de la
Investigaci\'on y Academia (CEDIA) through project
CEPRA-XIII-2019 ``Sistemas Complejos''.

\end{document}